\documentclass[letterpaper,12pt]{article}

\usepackage{amsmath}
\usepackage{amssymb}
\usepackage{setspace}

\newcommand{\eqn}[1]{Eqn. \eqref{#1}}
\newcommand{\eqns}[1]{Eqns. \eqref{#1}}
\newcommand{\Cite}[1]{Ref. \cite{#1}}
\newcommand{\ph}[1]{\phantom{#1}}

\newcommand{\Cal}[1]{\ensuremath{\mathcal{#1}}}
\newcommand{\Mbar}{\ensuremath{\bar{\Cal{M}}}}
\newcommand{\avg}[1]{\ensuremath{\left\langle #1\right\rangle}}

\newcommand{\rmb}[1]{{\bf #1}}

\newcommand{\ba}{\ensuremath{\bar a}}

\begin{document}
\begin{center}
{ \Large {\bf A Covariant Road to Spatial Averaging in Cosmology :}}\\
{ {\large\bf Scalar Corrections to the Cosmological Equations}}
\vskip 2cm
{\large\bf Aseem Paranjape\footnote{e-mail address: aseem@tifr.res.in}}
\vskip 1cm
{\it Tata Institute of Fundamental Research,}\\
{\it Homi Bhabha Road, Mumbai 400 005, India.} \vskip 1.5cm
This essay received an Honourable Mention in the Gravity Research
Foundation's Essay Competition, 2007.
\end{center}
\vskip 1cm
\begin{abstract}
\bigskip
\noindent 
A consistent approach to Cosmology requires an explicit averaging of
the Einstein equations, to describe a homogeneous and isotropic
geometry. Such an averaging will in general modify the Einstein
equations. The averaging procedure due to Buchert has attracted
considerable attention recently since it offers the tantalizing hope
of explaining the phenomenon of dark energy through such
corrections. This approach has been criticized, however, on the
grounds that its effects may be gauge artifacts. We apply the fully
covariant formalism of Zalaletdinov's Macroscopic Gravity and show
that, after making some essential gauge choices, the Cosmological
equations receive \emph{spacetime scalar} corrections which are
therefore observable in principle, and further, that the broad
structure of these corrections is \emph{identical} to those derived by 
Buchert.   
\end{abstract}
\newpage

\begin{doublespacing}
Cosmology crucially assumes that the matter distribution in the 
Universe, when averaged on large enough scales, is homogeneous and
isotropic. Assuming that the metric can also be ``averaged'' on these 
scales, one then models this averaged metric of spacetime as having 
the Friedmann-Lema\^itre-Robertson-Walker (FLRW) form with homogeneous 
and isotropic spatial sections, and applies Einstein's General 
Relativity to determine the geometry of the Universe. The averaging 
operation is usually assumed implicitly and rather vaguely. It has 
long been known however \cite{ellis}, that any \emph{explicit} 
averaging scheme for the metric $g$ of spacetime and energy-momentum 
tensor $T$ of matter, must necessarily yield corrections to the 
Einstein equations (which should ideally be imposed on length scales 
comparable to, say, the Solar System). This is a consequence of the 
Einstein tensor $E[g]$ being a nonlinear functional of the metric $g$ 
: given some averaging operator $\avg{\cdot}$, in general one has  
$\avg{E[g]} \neq E[\avg{g}]$, and by using $E[\avg{g}]=\avg{T}$ as the 
field equations, one is ignoring these corrections.

Clearly, one needs a systematic averaging scheme within which to 
ask how large these corrections are. Among the averaging schemes 
available in the literature, only two -- Buchert's spatial averaging 
of scalars \cite{buch} and Zalaletdinov's Macroscopic Gravity (MG) 
\cite{zala} -- are capable of addressing the issue of averaging in 
General Relativity in a \emph{nonperturbative} manner. (This is 
important since one expects the effects of averaging, if any, to show 
up only through nonlinear inhomogeneities.) Buchert's approach has  
attracted considerable attention recently due partly to its  
tractability, but mainly because it offers the tantalizing hope of 
solving the dark energy problem \emph{completely within the framework 
  of classical General Relativity}. However, this approach has been 
criticized on the grounds that an averaging operation such as 
Buchert's which is defined only for a particular $3+1$ splitting of 
spacetime, is likely to lead to observationally irrelevant gauge 
artifacts which could wrongly be interpreted as solving the dark
energy problem.

While it is difficult to refute such criticism from within Buchert's 
noncovariant approach, it might be possible to address the issue 
beginning with a \emph{covariant} averaging scheme. Zalaletdinov's MG 
is precisely such a scheme -- fully covariant and mathematically 
elegant -- which unfortunately comes with the somewhat steep price of 
extreme technical complexity. Nevertheless, the following question can 
be posed : Starting with the fully covariant structure of MG, and then 
making some appropriate gauge choices, is it possible to derive 
equations resembling Buchert's modified FLRW equations? We answer 
this question in the affirmative \cite{spatlim}, and address some of 
the issues that emerge in this construction. We begin by describing  
the broad structure of MG.

Zalaletdinov's approach considers a general differentiable manifold 
\Cal{M} with metric $g_{ab}$, and defines a spacetime averaging 
operation for tensors which is then used to \emph{construct} an  
``averaged'' differentiable manifold \Mbar. The average of the affine  
connection on \Cal{M}\ is shown to itself behave like a connection, 
and is used as the affine connection for the abstract manifold 
\Mbar. The averaging operation is very sophisticated -- it is defined 
using a Lie dragging of averaging regions along chosen vector fields, 
and ensures that the average of some tensor field $p{}^i_j(x)$, say, 
is itself a local tensor field $\avg{p{}^i_j}(x)$ on the manifold 
\Cal{M}. One defines the (tensorial) connection correlation terms, 
\begin{equation}
Z^{a\ph{bm}i}_{\ph{a}b[m\ph{i}\underline{j}n]} =
\avg{\Gamma{}^a_{b[m}
    \Gamma{}^i_{\underline{j}n]}} -
\avg{\Gamma{}^a_{b[m}}
  \avg{\Gamma{}^i_{\underline{j}n]}} ~~;~~
Z^a_{\ph{a}ijb} = 2 Z^{a\ph{ik}k}_{\ph{a}ik\ph{k}jb} \,,
\label{eq2}
\end{equation}                                                      
where the square brackets denote antisymmetrization and the underlined 
indices are not antisymmetrized. It can then be shown that the 
averaged Einstein equations read 
\begin{equation}
E{}^a_b = -\kappa T{}^a_b + C{}^a_b\,,
\label{eq3}
\end{equation}                                                    
where $\kappa=8\pi G_N$, $E{}^a_b$ is the Einstein tensor for \Mbar, 
$T{}^a_b$ is the averaged energy-momentum tensor and the correlation  
tensor $C{}^a_b$ is defined as
\begin{equation}
C{}^a_b = \left(Z^a_{\ph{a}ijb} - \frac{1}{2}\delta{}^a_b
Z^m_{\ph{a}ijm} \right)G^{ij}\,,
\label{eq4}
\end{equation}                                                    
where $G_{ab}=\avg{g_{ab}}$ is the metric on \Mbar, which we assume to 
be a highly symmetric space. The tensor $C{}^a_b$ also satisfies  
$C{}^a_{b;a} =0$, where the semicolon denotes covariant derivative on 
\Mbar.

We now make two important assumptions. Firstly, starting with some 
(formally unspecified) choice of $3+1$ splitting for the inhomogeneous
spacetime, we consider the \emph{time} averaging scale (denoted $T$)
to be much smaller than the typical scale of time evolution of the
metric. This is justified physically in the cosmological context since
all measurements are performed over timescales much smaller than the  
Hubble expansion scale. Formally then, we consider the $T\to0$, 
``spatial averaging limit'' of MG. (This limit is intuitively seen to 
be well defined for averages of smooth functions, and this can also be 
verified explicitly.) Secondly, we assume that the averaging 
\emph{length} scale (denoted $L$) is of order $\sim150$-$200$Mpc, the 
scale at which cosmological homogeneity sets in. The averaged manifold 
\Mbar\ is then assumed to admit a unit timelike vector field $\bar
v^a$,                                                                
which is orthogonal to spacelike hypersurfaces of constant    
curvature -- in other words, \Mbar\ is the FLRW spacetime. The spatial 
averaging limit is important since it ensures that the FLRW spacetime 
is a fixed point of the averaging scheme. Further treating the  
averaged energy-momentum tensor to have the form of a homogeneous and 
isotropic perfect fluid, one can write down two \emph{scalar}  
equations -- the (modified) Friedmann and Raychaudhuri equations -- 
given by
\begin{equation}
\left(\frac{1}{a}\frac{da}{d\tau}\right)^2 = \frac{8\pi G_N}{3}\rho  
-\frac{1}{3}C{}^a_b\bar v^b\bar v_a ~~;~~
\frac{1}{a}\frac{d^2a}{d\tau^2} = -\frac{4\pi
  G_N}{3}\left(\rho+3p\right) + \frac{1}{6}\left(C{}^a_a +
2C{}^a_b\bar v^b\bar v_a\right)\,. 
\label{eq5}
\end{equation}                                                 
We have assumed that the averaged metric takes the (flat, for   
simplicity) FLRW form in the synchronous gauge, with scale factor 
$a(\tau)$. The specific gauge chosen is irrelevant since these are
scalar equations. $\rho$ and $p$ are the homogeneous energy density
and pressure respectively as measured by ``observers'' in \Mbar\  
travelling on trajectories with tangent vector field $\bar
v^a$.                                                          
Physically these observers correspond to the \emph{averaging  
  domains} treated as physically infinitesimal cells -- this is a 
refinement of the Weyl postulate which treated individual galaxies as 
travelling on hypersurface orthogonal trajectories. It is obvious from 
\eqn{eq5} that the corrections to the standard cosmological equations 
arising from averaging of inhomogeneities, are \emph{spacetime  
  scalars}. In addition, we are forced to set the $3$-vector  
($C{}^0_A, A=1,2,3$) and traceless $3$-tensor ($C{}^A_B -
(1/3)\delta{}^A_B(C{}^J_J)$) parts of $C{}^a_b$ to zero, due to the 
structure of $E{}^a_b$ and the assumed form of $T{}^a_b$. This imposes 
additional consistency constraints on the inhomogeneous geometry.

The main question we wish to address concerns the \emph{structure} 
of the corrections. Immediately, this leads to another question, 
namely, what choice of gauge for the \emph{inhomogeneous} geometry,
will lead to the averaged metric in the standard synchronous FLRW 
form? This question has no clear answer, and it is perhaps wiser to 
simply leave this gauge choice formally unspecified. It turns out 
however, that the form of the corrections simplifies if one begins 
with a \emph{volume preserving gauge}, in which the metric determinant  
is constant. This happens because the structure of the MG formalism
itself becomes very simple in the volume preserving gauge
\cite{zala}. We will therefore display these results to highlight the
broad structure of these corrections. Let us suppose that the
inhomogeneous metric is written in coordinates wherein it takes the
form    
\begin{equation}
^{(\Cal{M})}ds^2=-\frac{dt^2}{h(t,\rmb{x})} +
  h_{AB}(t,\rmb{x})dx^Adx^B ~~;~~ A,B=1,2,3\,, 
\label{eq6}
\end{equation}                                                
with $h={\rm det}(h_{AB})$, so that $\sqrt{-g} =
(1/\sqrt{h})(\sqrt{h})=1$. If we now assume that the ``spatial'' 
average of $h_{AB}$ is  
\begin{equation}
\avg{h_{AB}}=\ba^2(t)\delta_{AB}\,,
\label{eq7}
\end{equation}                                               
then it can be shown \cite{spatlim} that the full averaged metric must 
take the volume preserving form
\begin{equation}
^{(\Mbar)}ds^2 = -\frac{dt^2}{\ba^6(t)} +
  \ba^2(t)\delta_{AB}dx^Adx^B \,,
\label{eq8}
\end{equation}                                           
which can be brought to synchronous form by the transformation 
$\tau=\int^t{dt/\ba^3}$. Further, writing $a(\tau)=\ba(t(\tau))$, the 
modified Cosmological equations now read
\begin{subequations}
\begin{align}
\left(\frac{1}{a}\frac{da}{d\tau}\right)^2 &= \frac{8\pi G_N}{3}\rho - 
  \frac{1}{6}\left[ \Cal{Q} + \Cal{S} \right] \,,
\label{eq9a} \\&\nonumber\\
\frac{1}{a}\frac{d^2a}{d\tau^2} &= -\frac{4\pi G_N}{3}\left(\rho + 
  3p\right) + \frac{1}{3}\Cal{Q} \,,
\label{eq9b} 
\end{align}
\label{eq9} 
\end{subequations}                                          
where  $\Cal{Q}$ and $\Cal{S}$ are given by 
\begin{subequations}
\begin{align}
\Cal{Q} &= \ba^6\left[\frac{2}{3}\left(\avg{\frac{1}{h}\Theta^2} -
\frac{1}{\ba^6}(^{\rm F}\Theta^2)\right) -
2\avg{\frac{1}{h}\sigma^2}\right] \,,
\label{eq10a} \\&\nonumber\\
\Cal{S} &= \frac{1}{\ba^2}\delta^{AB}\left[
  \avg{\,^{(3)}\Gamma{}^J_{AC}\,^{(3)}\Gamma{}^C_{BJ}} -
  \avg{\partial_A(\ln\sqrt{h})\partial_B(\ln\sqrt{h})}  
   \right] \,.
\label{eq10b}
\end{align}
\label{eq10}
\end{subequations}                                         
Here $\Theta$ and $\sigma^2$ are the expansion and shear scalars 
respectively for the inhomogeneous geometry in the gauge 
\eqref{eq6}, $\,^{(3)}\Gamma{}^A_{BC}$ is the Christoffel symbol 
constructed using $h_{AB}$, and the expansion scalar for the 
\emph{averaged} geometry in the volume preserving gauge is given by 
$(1/\ba^3)(^{\rm F}\Theta) = (3/\ba)(d\ba/dt)$. $\Cal{Q}$ and 
$\Cal{S}$ are defined in the volume preserving gauge, but they are 
considered functions of the synchronous time $\tau$ in \eqns{eq9}, 
satisfying the condition  
\begin{equation}
\left(\partial_\tau\Cal{Q} + 6\Cal{Q}
 \partial_\tau(\ln a) \right) +  \left(\partial_\tau\Cal{S} +  
  2\Cal{S}\partial_\tau(\ln a)\right) =  0 \,,
\label{eq-deriv}
\end{equation}                                            
which follows from $C{}^a_{b;a}=0$. The additional consistency 
conditions mentioned earlier read 
\begin{subequations}
\begin{align}
&\delta^{JK}\left[\avg{\sqrt{h}\Theta_{JB}\,^{(3)}\Gamma{}^B_{AK}} -
  \avg{\sqrt{h}\Theta_{JK}\,^{(3)}\Gamma{}^B_{AB}} \right] = 0\,,  
\label{eq11a}\\
&\delta^{JK}\avg{\frac{1}{\sqrt h}\Theta{}^B_K\,^{(3)}\Gamma{}^A_{JB}}
  - \delta^{AJ}\avg{\frac{1}{\sqrt
      h}\Theta\,^{(3)}\Gamma{}^B_{JB}} = 0\,,  
\label{eq11b}\\
&\delta^{JK}\avg{\,^{(3)}\Gamma{}^A_{JC}\,^{(3)}\Gamma{}^C_{KB}} -
\delta^{AJ}\avg{\,^{(3)}\Gamma{}^C_{JC}\,^{(3)}\Gamma{}^K_{BK}} = 
\frac{1}{3}\delta{}^A_B\left(\ba^2\Cal{S}\right) \,,
\label{eq11c}
\end{align}
\label{eq11}
\end{subequations}       
with $\Theta{}^A_B$ the expansion tensor.

Our result contains some striking features. Firstly, despite the fact 
that the MG approach is very different from Buchert's spatial 
averaging (where, e.g., scalar equations are constructed \emph{before} 
averaging as opposed to our \eqns{eq5}, see also \Cite{spatlim}), our 
final results in \eqns{eq9} and \eqref{eq-deriv} are formally 
\emph{identical} in structure to Buchert's modified Cosmological 
equations (see, e.g., Eqns. (10) and (13) in the first of 
\Cite{buch}). We can therefore legitimately claim that corrections 
similar to those derived by Buchert arise as \emph{spacetime scalars} 
in the covariant MG approach. Further, \eqns{eq11} indicate that it is 
not sufficient to demand that an inhomogeneous metric average out to 
the FLRW form -- additional correlations are also required to 
vanish and this imposes further conditions on the inhomogeneous
metric.

To conclude, we have argued in the framework of Zalaletdinov's 
covariant MG, that the modifications to the standard FLRW equations 
are spacetime scalars, and hence in principle, observables. Moreover, 
under some simplifying assumptions we have demonstrated that the 
\emph{structure} of these corrections, and the equations they satisfy, 
are formally \emph{identical} to that of similar corrections derived 
by Buchert via a completely different approach. The question of 
whether these corrections can account for the observed acceleration of
FLRW models, while still open, is now on a concrete, observationally
relevant footing.

\end{doublespacing}

\end{document}